\documentclass[12pt,preprint]{aastex}

\usepackage{graphicx}

\newcommand{\lya}{Ly-$\alpha$\ }

\def\mb{\ifmmode {{\rm B_{435}}}\else
                ${\rm B_{435}}$\fi}
\def\mv{\ifmmode {{\rm V_{606}}}\else
                ${\rm V_{606}}$\fi}
\def\mi{\ifmmode {{\rm i_{775}}}\else
                ${\rm i_{775}}$\fi}
\def\mz{\ifmmode {{\rm z_{850}}}\else
                ${\rm z_{850}}$\fi}
\def\mJ{\ifmmode {{\rm J_{1100}}}\else
                ${\rm J_{1100}}$\fi}
\def\mH{\ifmmode {{\rm H_{1600}}}\else
                ${\rm H_{1600}}$\fi}
                
\def\mJi{\ifmmode {{\rm Ji_{1100}}}\else
                ${\rm Ji_{1100}}$\fi}
\def\mHi{\ifmmode {{\rm Hi_{1600}}}\else
                ${\rm Hi_{1600}}$\fi}
\def\mKs{\ifmmode {{\rm Ks_{2200$\ \AA$}}}\else
                ${\rm Ks_{2200\ \AA}}$\fi}

\def\mIRACa{\ifmmode {{\rm IRAC_{3600}}}\else
                ${\rm IRAC_{3600\AA}}$\fi}
\def\mIRACb{\ifmmode {{\rm IRAC_{4500}}}\else
                ${\rm IRAC_{4000\AA}}$\fi}
                
\def\ergcm2s{\ifmmode {\rm\,erg\,cm^{-2}\,s^{-1}}\else
                ${\rm\,ergs\,cm^{-2}\,s^{-1}}$\fi}
   
\newcommand{\nlya}{9\ }
             
\newcommand{\mpcv}{\rm \ Mpc^{-3}}

\newcommand{\ferg}{\rm \ ergs\ s^{-1}\ cm^{-2}}
 
\newcommand{\Prob}{$P({\chi}_{{\nu}c}^2 > {{\chi}_{{\nu}c_0}^2})$}
 
\newcommand{\Zunit}{${\rm  \% Z_{\sun}}$}

\begin{document}
\title{Optical-to-Mid-Infrared Observations of Ly-$\alpha$ Galaxies at z$\approx$5 in the Hubble Ultra Deep Field: A Young and Low Mass Population}

\author{N. Pirzkal\altaffilmark{1,2},S. Malhotra\altaffilmark{3},J. E. Rhoads\altaffilmark{3},C. Xu\altaffilmark{4}
}
\altaffiltext{1}{Space Telescope Science Institute, Baltimore, MD 21218.}
\altaffiltext{2}{Affiliated with the Space Science Depatment of the European Space Agency, ESTEC, NL-2200 AG Noordwijk,  Netherlands.}
\altaffiltext{3}{School of Earth and Space Exploration, Arizona State University, Tempe, AZ 85287.}
\altaffiltext{4}{Shanghai Institute of Technical Physics, 200083, Shanghai, China.}


\begin{abstract}
High redshift galaxies selected on the basis of strong
Ly-$\alpha$ emission tend to be young and have small physical
sizes. We show this by analyzing the spectral energy distribution of nine
Ly-$\alpha$-emitting galaxies (LAEs)  at $4.0 < z < 5.7$ in the Hubble Ultra Deep Field
. Rest-frame UV to optical $700{\rm \AA} < \lambda < 7500 {\rm
\AA}$ luminosities, or upper limits, are used to constrain old stellar
populations. We derive best fit, as well as maximally massive and
maximally old, properties of all nine objects. We show that these faint and distant 
objects are all very young, most likely only a few million
years old, and not massive, the mass in stars being $\approx 10^6-10^8
{\rm \ M_{\sun}}$. Deep Spitzer Infrared Array Camera observations of
these objects, even in cases where the object was not detected, proved
crucial in constraining the masses of these objects. The space density
of these objects, $\approx 1.25 \times 10^{-4} \mpcv$, is comparable to
previously reported space densities of LAEs at moderate-to-high
redshifts. These Ly-$\alpha$ galaxies show modest star formation
rates of $\approx 8 {\rm \ M_{\sun}\ yr^{-1}}$, which is nevertheless
strong enough to have allowed them to assemble their stellar
mass in less than a few million years. These sources
appear to have small physical sizes, usually smaller than ${\rm 1\
kpc}$, and are also rather concentrated. They are likely to be some of
the least massive and youngest high redshift galaxies observed to
date.
\end{abstract}

\section{Introduction}
The availability of the Infrared Array Camera (IRAC) of the Spitzer Space Telescope has recently made it possible to measure the rest-frame optical light from  high-redshift galaxies, up to redshifts of $\approx 6$ \citep[e.g., ][]{yan2006,chary2005,egami2005,eyles2005,schaerer2005,dow2005,stark2007}. Indeed, in just the past few years large numbers of Lyman break galaxies (LBGs) and Ly-$\alpha$-emitting galaxies (LAEs) have been studied, and IRAC data have enabled mass and age estimates for these objects to be derived. Historically, LBGs and LAEs are two distinct classes of objects that have been observed over a very wide range of redshifts, thanks in large part to the fact that they are relatively straight-forward to identify by virtue of the presence of either a significant break in their spectral light distribution (LBGs) or the presence of a strong Ly-$\alpha$ emission line (LAEs). As the number of known $z > 5$ LBGs and LAEs increased over the past few years \citep[; references therein]{dawson2004, eyles2005, yan2006}, the mass estimates of these objects have remained somewhat uncertain. Nevertheless observations have shown that some high redshift LBGs tend to already be quite massive ($> 10^{10} {\rm \ M_{\sun}}$) and to be a few hundred million years old. The relatively large masses of these objects at a time when the universe was still quite young has been taken as an indication that  these objects have had a star formation history (SFH) that allowed for most of their stellar mass to have formed early on. 

Mass estimates for LAEs, on the other hand, have been much more difficult to assess. Either the LAEs are very young, as indicated by the strong \lya\ line emission or the faintness 
of the continuum, for some other reason (e.g., extinction) makes it difficult to derive mass estimates using their spectral energy distribution (SED) as is done for LBGs.  \citet{gawiser2006}, for example, required the use of a stack of 18 field LAEs to infer an average mass of $\approx 5 \times 10^{8} {\rm \ M_{\sun}}$.  This low estimated mass, and the low luminosity of these objects in the rest-frame optical, is exactly what makes estimating their masses somewhat more difficult than in the case of LBGs. While the latter are usually selected based on broad band observations and colors, hence ensuring that these objects have bright continuum luminosities, LAEs selected based on the presence of \lya emission can be significantly fainter in the observed bands.
For the handful of LAEs for which detections by IRAC have been obtained, a large mass in older populations has been derived \citep{lai2006,chary2005}. There is however a {\it detection bias} here: the only published mass/age estimates so far are the ones for which IRAC detections exist; that is, they would tend to have significant old stellar populations.

In this paper, we investigate the population of high redshift LAEs in the  Hubble Ultra Deep Field \citep[HUDF;][]{beckwith2007}, a field that is relatively small (11 ${\rm arcmin^2}$) but has the advantage of having been observed down to very faint magnitudes, and in a wide range of wavelengths. Our LAE sample makes no {\it apriori} assumptions as to the specific redshift of these sources, or as to whether these objects have been detected in all of the available data. We therefore aimed at reaching down to potentially fainter objects, and over a wider range of redshift, than previously done. Section \ref{observations} describes the object selection and the available data. In Section \ref{phot}, we summarize the broad-band luminosities of these objects, while in Sections \ref{Morphology}, \ref{popsynth}, and \ref{ages} we discuss the morphology and SED-fitting process and provide mass and age estimates of the HUDF LAEs, irrespective of whether these objects were detected in the Spitzer IRAC bands or not. 

\section{Observations}\label{observations}
\begin{figure}[h]
\includegraphics[width=7.0in]{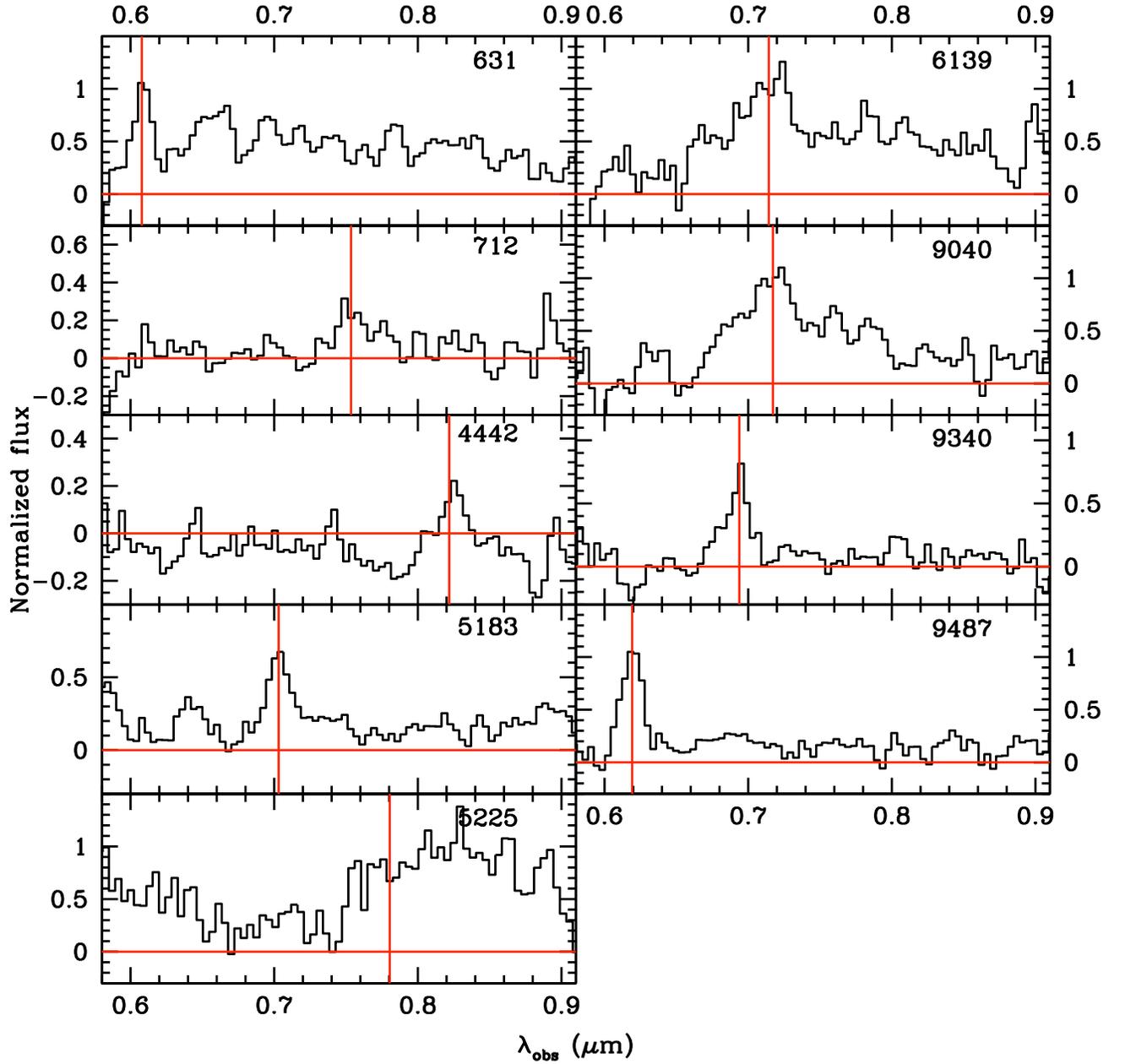}
\caption{\label{specs} ACS G800L grism spectra of the nine \lya\ galaxies
in our sample.}
\end{figure}

The sources described in this paper were selected from the GRAPES survey \citep[GRism ACS Program for Extragalactic Science, PI: S. Malhotra; see description in ][]{pirzkal2004}, a slitless grism spectroscopic  survey using the Advanced Camera for Surveys (ACS) on the Hubble Space Telescope (HST). The GRAPES project obtained low resolution spectra of 1421 objects in the HUDF down to a continuum level of $\mz = 27$ mag.  The GRAPES grism observations proved able to detect emission lines with fluxes as low as $\sim 5\times 10^{-18}\ \ferg$ in objects as faint as $\mz = 29$ mag in the continuum. A complete list of GRAPES emission line objects was published in \citet{xu2007} and contains 115 objects with redshifts $0.1 < z < 5.7$. GRAPES identified a total of \nlya LAEs at $z > 4.0$ (listed in Table \ref{objecttable}). We show individual spectra of these nine sources in Figure \ref{specs}. In most cases, the spectra shown are a narrow extraction of the GRAPES program data.  In two cases, where the galaxy is comparatively large, we show a wider extraction (5225 and 6139).  Finally, in another two cases (631 and 9040), the
GRAPES data have relatively few uncontaminated roll angles, and we therefore instead plot spectra from the subsequent PEARS program (Probing Evolution and Reionization Spectroscopically; PI: S. Malhotra), which reobserved the HUDF in four new roll angles. The emission lines in the largest objects (6139, 9040, and especially 5225) are not shown to best advantage in these one-dimensional extractions, because the spatial extent of the object gives a wide line-spread function in the slitless data, which tends to blur the line. Detailed two dimensional images of the observations of object 5225 are available in \citet{rhoads2005}. 
The GRAPES emission line object selection was performed automatically using
an emission-line search code and methodology described in details in \citet{xu2007}. 

Some of the objects shown in Table \ref{objecttable}  have rather faint emission lines.
All lines were nevertheless significant at the level of ${\rm 3.1 (N/3)^{1/2} \sigma}$,
where $N$ is the number of GRAPES position angles in which the line was detected,
and where $N\ge 2$ was required for inclusion in \citet{xu2006}.  For an emission
line object to be classified as a \lya\ galaxy, it had to also have (1) no
other emission lines in the G800L grism wavelength coverage\footnote{The case
of \lya\ plus CIV $\lambda$1549 emission did not occur in the GRAPES data set.}
and (2) broad band photometry in the HUDFimages consistent
with a Lyman break at the redshift corresponding to the \lya\ line
identification.  Given less accurate photometry, a \lya\ line plus Lyman
break may be difficult to distinguish from an [OII] 3727\AA\ line plus 4000\AA\ break
\citep{stern2000}.  Fortunately, even the faintest object in our sample is detected
witha signal-to-noise (S/N) greater than 5 two filters, and the rest are detected with (S/N)$> 15$ in two or more
filters.  This deep photometry shows that every object has a break where the flux is
attenuated by a factor of at least 15 (see Table \ref{phottable}).
The strengths of these breaks are several times greater than
would be expected from the D(4000) break, and are much
more consistent with the decrement expected from absorption by the intergalactic medium (IGM)
for objects at $z\approx 5$ \citep{madau1995}.  Also, although it was not
part of the GRAPES LAE selection process, these objects typically
have very blue observed $z-J$ colors (corresponding to
$\sim 1400-1800{\rm \AA}$ rest-frame near-UV colors), based on the
Near-Infrared Camera and Multi-Object Spectrometer (NICMOS) HUDF observations \citep{thompson2005}.
This makes it highly unlikely that any large amount of extinction is
present in these objects and so further supports the interpretation of
their colors as a Lyman break at high redshift.
Further spectroscopy could certainly help secure the redshifts
and \lya\ line identification for these objects.
For example, higher resolution ($R\ga 1000$) could help by verifying
that the lines show the asymmetry characteristic of \lya\ \citep[e.g.][]{rhoads2003,dawson2004}.  Similarly, bluer optical coverage
could confirm the absence of other lines, and infrared coverage
could confirm the presence of [OII] $\lambda3727$\AA\  at the same redshift as
\lya.  However, all present evidence points to these objects
being \lya\ sources at $z\approx 5.5$.
\\
The HUDF, which lies within the Great Observatories Origins Deep Survey \citep[GOODS;][]{giavalisco2004} field, has the advantage of having been observed in a wide range of wavelengths. In this paper, we take advantage of the availability of HST NICMOS, VLT ISAAC, and Spitzer IRAC observations of this field, as well as the list of nine spectroscopically identified GRAPES LAEs. At the redshifts considered in this paper ($4.0 < z < 5.7$), combining these observations together allows us to examine these objects at rest-frame wavelengths of approximately $700{\rm \AA} < \lambda < 7500 {\rm \AA}$, and down to unprecedented depths. The ACS observations, using the \mb, \mv, \mi, and \mz\ bands each reach down to approximately 29 mag, with a final mosaic image resolution of 0.030''  ${\rm pixel}^{-1}$ \citep{beckwith2007}. NICMOS was used to observe a large sub-set of the HUDF in the near infrared \citep{thompson2005} using both the J and H bands.  While the resolution of the NICMOS observations ($\approx 0.2''$ ${\rm pixel}^{-1}$) is  lower than that of the ACS observations, it is nevertheless high enough  to clearly separate LAEs from their neighbors. We additionally used the ISAAC (Infrared Spectrometer and Array Camera) Ks-band observations of the HUDF, but the depth of these data is relatively shallow (25.1 mag) and with a resolution  limited by atmospheric seeing ($\approx 0.5$''). Lastly, the HUDF was also observed in the infrared using IRAC as part of the GOODS Legacy program (M. Dickinson et al. 2007, in preparation). While the latter observations have the relatively low resolution of 1.2'' ${\rm pixel}^{-1}$, they offer a unique opportunity to probe the rest-frame optical properties of $z\approx 5$ LAEs. The IRAC data reach down to a depth of nearly 26 mag in both IRAC channel 1 ($3.6{\ \micron}$) and channel 2 ($4.5{\ \micron}$).

\begin{deluxetable}{rcccccccc}
\tabletypesize{\footnotesize} 
\tablewidth{0pc}
\tablecaption{GRAPES LAEs and Their Morphological Measurements. \label{objecttable}}
\tablecomments{The half-light radius ($R_{50}$), Concentration (C), Asymmetry (A) and absolute magnitude ($M_{1500\AA}$) were estimated at a rest-frame wavelength of $1500 {\rm \AA}$. The redshifts and the line fluxes are from \citet{xu2007}; the line fluxes from \citet{xu2007} shown in parenthesis were corrected using an aperture correction that we derived taking into account the size of each individual source. Units of right ascension are hours, miniutes, and seconds, and units of declination are degrees, arcminutes, and arcseconds.}
\tablenotetext{a}{\citet{rhoads2005}}
\tablehead{\colhead{UID} & \colhead{R.A.} & \colhead{Dec.} & \colhead{Redshift} &  \colhead{Flux}  & \colhead{$M_{1500\AA}$} &  \colhead{$R_{50}$} &  \colhead{C} &  \colhead{A}  \\
\colhead{} & \colhead{(J2000)} & \colhead{(J2000)} & \colhead{z} &  \colhead{$10^{-18} \ferg$}  & \colhead{} &  \colhead{(kpc)} &  \colhead{} &  \colhead{}
}
\startdata  
%631 & 3:32:40.09 & -27:49:1.19  & 4.127 & 20 & -20.21 & 0.67 & 2.74 & 0.097 \\
%712 & 3:32:42.81 & -27:48:58.51  & 5.1982 & 6 & -21.31 &  0.52 & 2.53 & 0.11 \\
%4442 & 3:32:39.38 & -27:47:39.17  & 5.7607 & 7 & -19.54 & 0.61 & 2.89 & 0.22 \\
%5183 & 3:32:34.51 &-27:47:27.11 & 4.7837 & 11 & -19.82 & 0.54 & 2.84 & $<0.1$ \\
%5225 & 3:32:33.26 & -27:47:24.76 & 5.524 & 2 & -22.77 & 1.86 & 2.18 & 0.31 \\
%6139 & 3:32:37.95 &-27:47:10.99 & 4.8765 & 22 & -22.14 & 1.35 & 3.19 & 0.41 \\
%%6515 & 3:32:30.57 & --27:47:6.60 & 4.7811 & 8 &  -20.77 & 0.57 & 2.81 & $<0.1$ \\
%9040 & 3:32:41.08 & --27:46:42.45 & 5.0452 & 13 &  -22.41 & 1.47 & 2.65 & 0.12 \\
%9340 & 3:32:40.67 & -27:45:56.11 & 4.7076 & 15 & -20.52 & 0.66 & 2.69 & 0.89 \\
%9487 & 3:32:40.17 & -27:46:0.55 & 4.0938 & 17 &  -21.98 & 0.62 & 2.36 & 0.11 \\
 631 & 3:32:40.09 & -27:49:01.19  & 4.00 & 58. (20) & -20.21 & 0.67 & 2.74 & 0.097 \\
 712 & 3:32:42.81 & -27:48:58.51  & 5.20 & 17. (6) & -21.31 &  0.52 & 2.53 & 0.11 \\
4442 & 3:32:39.38 & -27:47:39.17  & 5.76 & 24. (7) & -19.54 & 0.61 & 2.89 & 0.22 \\
5183 & 3:32:34.51 &-27:47:27.11 & 4.78 & 32. (11) & -19.82 & 0.54 & 2.84 & $<0.1$ \\
5225 & 3:32:33.26 & -27:47:24.76 & 5.42 & 22.\tablenotemark{a}& -22.77 & 1.86 & 2.18 & 0.31 \\
6139 & 3:32:37.95 &-27:47:10.99 & 4.88 & 60. (22) & -22.14 & 1.35 & 3.19 & 0.41 \\
%6515 & 3:32:30.57 & --27:47:6.60 & 4.7811 & 8 &  -20.77 & 0.57 & 2.81 & $<0.1$ \\
9040 & 3:32:41.08 & --27:46:42.45 & 4.90 & 36. (13) &  -22.41 & 1.47 & 2.65 & 0.12 \\
9340 & 3:32:40.67 & -27:45:56.11 & 4.71 & 49. (15) & -20.52 & 0.66 & 2.69 & 0.89 \\
9487 & 3:32:40.17 & -27:46:00.55 & 4.10 & 55. (17) &  -21.98 & 0.62 & 2.36 & 0.11 \\
\enddata
\end{deluxetable}

\section{Photometry of \lya emitters}\label{phot}
\begin{figure}[h]
\includegraphics[width=7.0in]{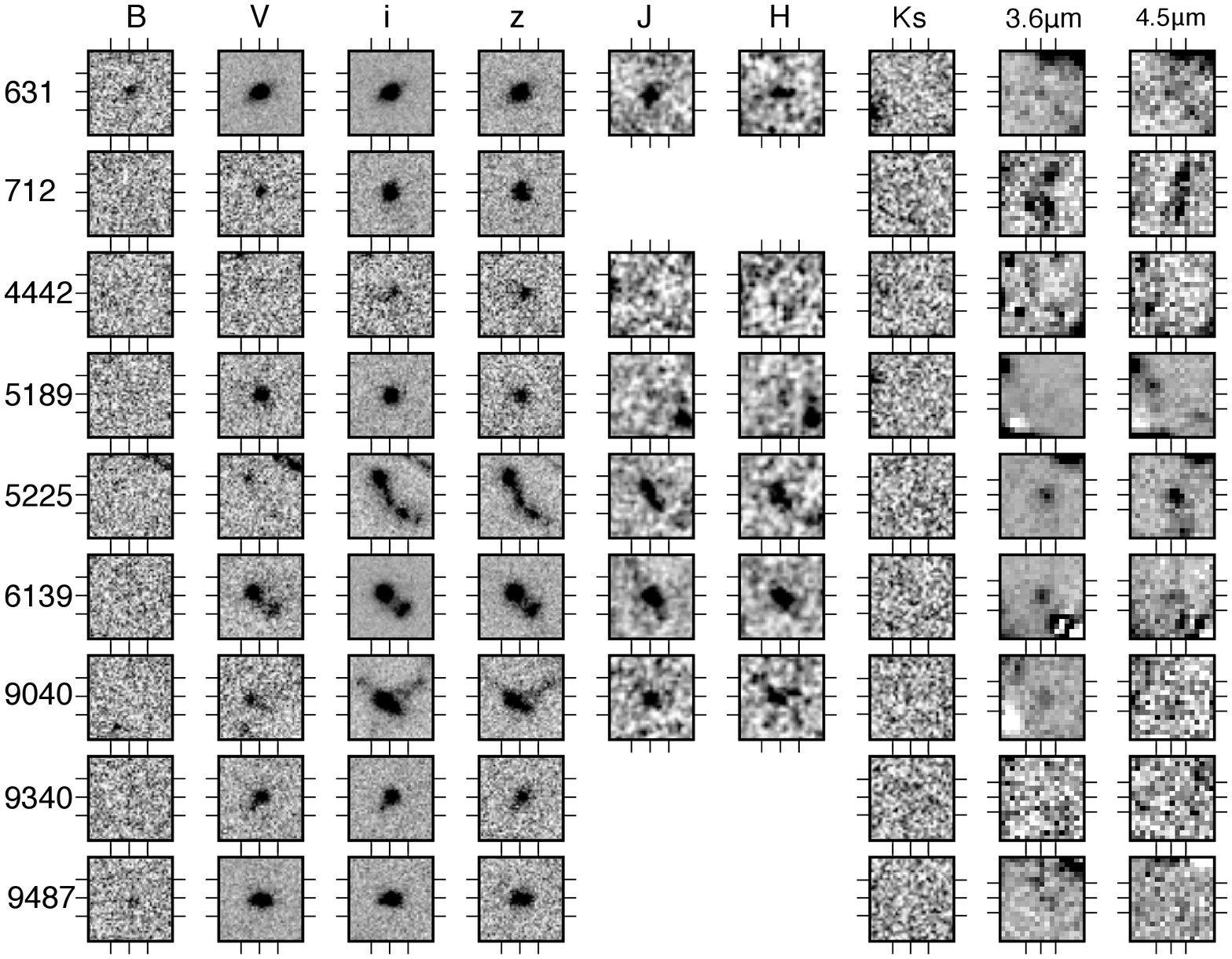}
\caption{\label{stps} ACS, NICMOS, ISAAC and IRAC observations of GRAPES LAEs. The source is at the center of each postage-stamp image. ACS \mb, \mv, \mi\ and \mz\ images are 1.2''-1.2'', NICMOS \mJ\ and \mH\ band images are 2.4''-2.4'', and ISAAC \mKs, \mIRACa\ and \mIRACb\ images are 5''-5''.}
\end{figure}

The luminosity of each source was measured separately in each available band. Postage stamp images of the GRAPES LAEs are shown in Figure \ref{stps}. Flux estimates were obtained using simple aperture photometry so that we could set meaningful upper limits to the flux of undetected sources, while the local noise level was estimated using a small annulus surrounding each source.  How we specifically dealt with non detections is explained in more detail in Section \ref{popsynth}.  We used nominally sized apertures, each containing as much of the object flux as possible while minimizing the flux contributions from nearby objects, by additionally using an object catalog derived using the ACS $\mi and \mz$ data. The latter allowed us to verify that contamination from neighboring sources remained small.\\
Different aperture sizes were used in different bands. With the ACS data, we used an aperture size set to 5 times the size of the object as measured by SExtractor \citep{bertin1996}  (in this paper, ``aperture size'' refers to the diameter of the aperture), and no aperture correction was required. We compared these measured fluxes with SExtractor measurements and found that the two agreed very well, to better than 1\%. We used an aperture of 1.25'' with the NICMOS data and applied aperture corrections of 1.06 and 1.09 in the J and H bands, respectively (NICMOS Support 2007, private communication). When measuring fluxes from the ground based ISAAC data, we used an aperture size that was twice as large as the size of the source, as measured in the ACS z-band image and after accounting for a seeing disk of  0.5''. As discussed above, the IRAC data both have a lower resolution and suffer from a higher level of contamination from neighboring sources than the rest of our data. The IRAC photometric measurement process was therefore more involved than when using ACS, NICMOS, and ISAAC data. Since many of the GRAPES LAEs were only marginally resolved in the IRAC observations, and because contaminating flux from nearby objects would cause flux measurements to be systematically higher (leading to systematically higher masses, as we discuss further below), we developed a technique to remove as much of the contaminating flux as possible.  Our approach consisted of performing two-dimensional general S\'ersic profile fits of all neighboring sources (whose positions were derived using the ACS $\mi and \mz$ object catalog) using the program GALFIT \citep{peng2002}. The centermost 4'' region surrounding each LAE was masked and therefore not used during the fitting process. Each object was individually and interactively processed until the best possible fit was achieved, while avoiding over-subtraction of the flux contribution from the neighboring sources. Following this, the two dimensional  models of the light profiles of nearby sources were subtracted from the original IRAC images and aperture photometry was performed on the resultant images, just as we did with the ACS, NICMOS, and ISAAC data. 

The reduced $\chi^2$ of the GALFIT fitting of the nearby sources was usually near unity. In the few cases where nearby sources were not perfectly fitted using a S\'ersic profile, we made sure that we did not end up over-subracting the amount of light contributed by these nearby objects.  At the presumed redshifts of these objects, undersubtracting the contaminating light produced by nearby object ultimately results in an overestimate of the rest-frame optical flux of the GRAPES LAEs. This, in turn, can only lead to an inflation of the stellar mass estimates. We used a fixed aperture of 3'' for all IRAC data and applied aperture corrections of 1.78 and 1.77 to IRAC channel 1 and channel 2, respectively  (B. Mobasher 2006, private communication).  The IRAC aperture corrections were determined following extensive Monte Carlo simulations \citep{mobasher2005}. Note that we did not use IRAC channels 3 and 4, as these proved a bit too noisy for our purpose.
Table \ref{phottable} shows the magnitude estimates for each source and in each available band, as well as the $3\ \sigma$ upper limit estimates for non detection.

\begin{deluxetable}{rrrrrrrrrrrrrr}
\tabletypesize{\footnotesize} 
\tablewidth{0pc}
\tablecaption{GRAPES LAEs Photometric Measurements}
\tablecomments{Error estimates are shown in parentheses; $3\ \sigma$ upper limits are indicated with greater-than signs.}
\rotate
\tablehead{

\colhead{} &  \multicolumn{4}{c}{ACS}  & \colhead{} &  \multicolumn{2}{c}{NICMOS}  & \colhead{} &  \multicolumn{1}{c}{ISAAC}  & \colhead{} &  \multicolumn{2}{c}{IRAC}  \\
\cline{2-5} 
\cline{7-8}
\cline{10-10}
\cline{12-13}
\colhead{UID}  & \colhead{\mb} &  \colhead{\mv} &  \colhead{\mi} &  \colhead{\mz} & \colhead{} &  \colhead{\mJ} &  \colhead{\mH} & \colhead{} & \colhead{\mKs} & \colhead{} &  \colhead{\mIRACa} &  \colhead{\mIRACb}}
\label{phottable}
\startdata  
631  &  29.59  &  26.87  &  26.41  &  26.47  & &  26.49  &  26.76  & &   $>26.61$  & &  27.17  &  27.16  \\
 &  (0.35)  &  (0.02)  &  (0.01)  &  (0.03)  & &  (0.07)  &  (0.10)   & &  (0.83)  & & (0.37)  &  (0.44)  \\
712  &  31.48  &  29.66  &  27.25  &  27.12  & &     &     & &   26.74  & &  28.05  &  $>27.45$  \\
  &  (2.01)  &  (0.25)  &  (0.03)  &  (0.04)  & &     &      & &  (1.32)  & & (0.84)  &  (0.83)  \\
4442  &  $>32.96$  &  31.64  &  29.35  &  28.36  & &  29.48  &  30.63  & &   $>27.78$  & &  $>28.53$  &  $>28.07$  \\
   &  (0.83)  &  (1.22)  &  (0.18)  &  (0.11)  & &  (0.52)  &  (2.04)   & &  (0.83)  & & (0.83)  &  (0.83)  \\
5183  &  31.76  &  28.27  &  27.39  &  27.97  & &  28.60  &  28.11  & &   $>27.33$  & &  $>28.13$  &  $>27.85$  \\
 &  (2.04)  &  (0.06)  &  (0.03)  &  (0.09)  & &  (0.28)  &  (0.20)   & &  (0.83)  & & (0.83)  &  (0.83)  \\
5225  &  28.88  &  28.29  &  25.97  &  25.93  & &  26.05  &  26.68  & &   26.54  & &  26.51  &  25.86  \\
  &  (0.41)  &  (0.17)  &  (0.02)  &  (0.04)  & &  (0.07)  &  (0.16)   & &  (1.62)  & & (0.28)  &  (0.26)  \\
6139  &  $>32.92$  &  26.93  &  25.52  &  25.66  & &  25.83  &  25.84  & &   26.20  & &  26.35  &  27.10  \\
  &  (0.83)  &  (0.04)  &  (0.01)  &  (0.02)  & &  (0.12)  &  (0.15)   & &  (0.90)  & & (0.23)  &  (0.39)  \\
%6515  &  $>33.00$  &  28.40  &  27.13  &  27.43  & &     &     & &   26.30  & &  28.73  &  $>28.07$  \\
 %  &  (0.83)  &  (0.07)  &  (0.03)  &  (0.05)  & &     &      & &  (0.83)  & & (1.05)  &  (0.83)  \\
9040  &  30.14  &  28.30  &  25.82  &  26.23  & &  26.16  &  26.11  & &   27.06  & &  $>28.45$  &  $>28.36$  \\
 &  (1.29)  &  (0.17)  &  (0.02)  &  (0.05)  & &  (0.14)  &  (0.22)   & &  (2.49)  & & (0.83)  &  (0.83)  \\
9340  &  33.04  &  28.09  &  27.51  &  27.69  & &     &     & &   $>27.40$  & &  $>27.48$  &  $>27.16$  \\
 &  (0.83)  &  (0.05)  &  (0.03)  &  (0.07)  & &     &      & &  (0.83)  & & (0.83)  &  (0.83)  \\
9487  &  30.07  &  27.25  &  27.16  &  27.27  & &     &     & &   $>27.24$  & &  27.47  &  $>27.93$  \\
   &  (0.47)  &  (0.02)  &  (0.03)  &  (0.05)  & &     &      & &  (0.83)  & & (0.47)  &  (0.83)  \\
%9999  &  30.61  &  31.02  &  30.88  &  30.26  & &  27.02  &  24.94  & &   23.95  & &  22.09  &  21.80  \\
%  &  (0.30)  &  (0.30)  &  (0.30)  &  (0.30)  & &  (0.32)  &  (0.07)   & &  (0.13)  & & (0.10)  &  (0.10)  \\

\enddata
\end{deluxetable}

\section{Morphology}\label{Morphology}
\begin{figure}[h]
\includegraphics[width=7.0in]{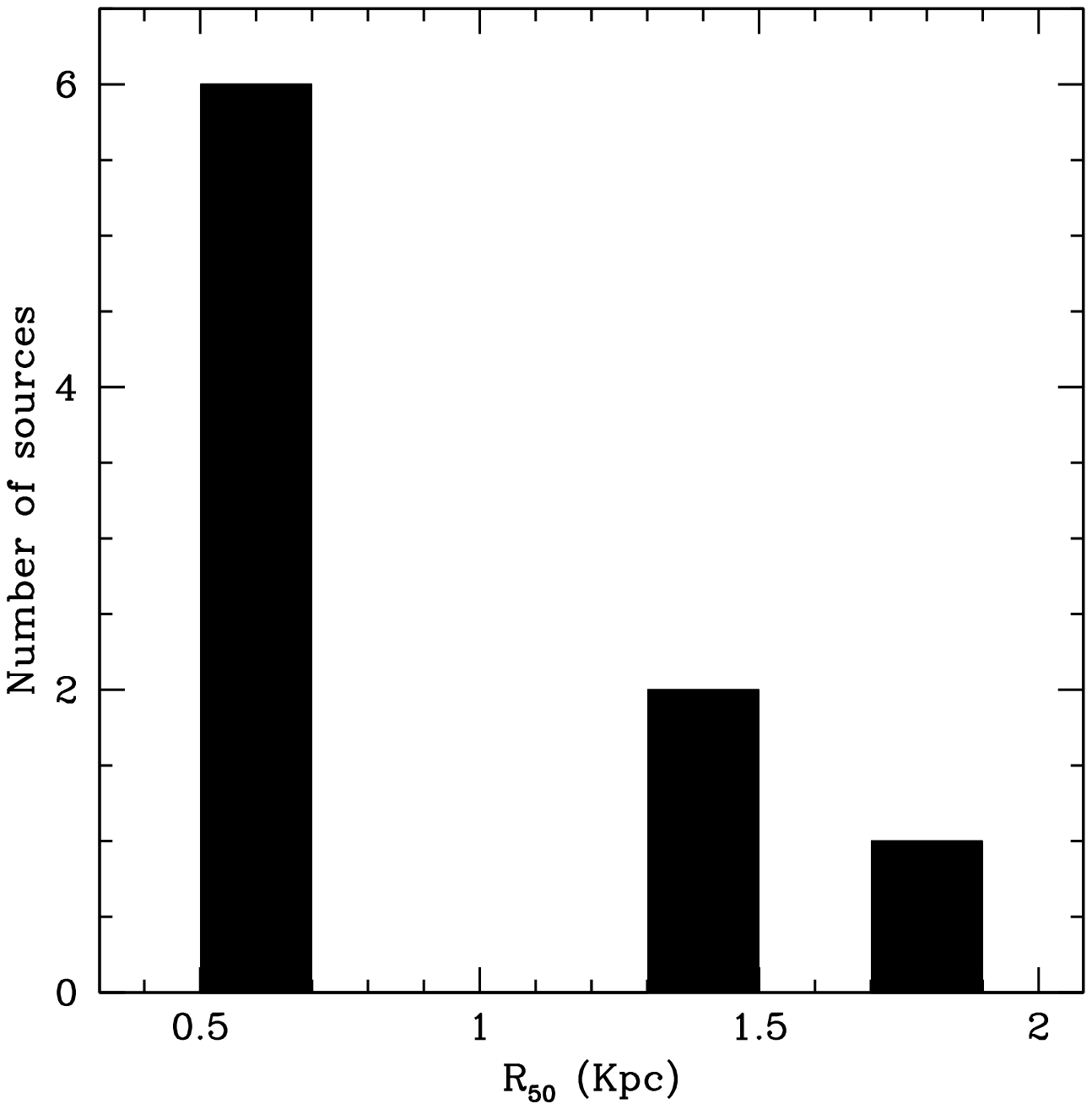}
\caption{\label{sizehist} Histogram of the measured half-light radius (${\rm R_{50}}$) of the GRAPES LAEs. Most objects are observed to have small radii, with ${\rm R_{50} < 1.0 \ kpc}$.}
\end{figure}

We measured the size and morphology of each GRAPES LAE using both GALFIT \citep{peng2002} and the CAS parameters \citep{conselice2000}. We computed the rest-frame size, concentration (C) and asymmetry (A) of each source separately in each of the available ACS bands. $1500{\rm \ \AA}$ rest-frame values of these parameters were then computed by interpolating these measurements. \\
The individual measurements are shown in Table \ref{objecttable}. GRAPES LAEs are found to be sharply concentrated and asymmetric sources, with average values of $2.67 \pm 0.30$ and $0.23 \pm 0.30$ for C and A, respectively. The half-light radii obtained using GALFIT are consistent with this, with an average size of  $0.92 \pm 0.50 $ kpc (or $0.17 \pm 0.07 {\rm ''}$) while best-fitting light profiles were sharp and concentrated in most cases. Three of the sources (5225, 6139, and 9040) are significantly larger than the rest of the GRAPES LAEs, however, with  an average half light radius of $1.56 \pm 0.27 {\rm \ kpc}$, as shown in Figure \ref{sizehist}. Moreover, two of these three objects (5225 and 6139)  have larger measured asymmetry values than the rest of the sample. In general, objects with larger asymmetries are observed to be have large half-light radii, raising the possibility that these objects could in fact be the result of smaller interacting components. The only exception is object 9340 which, while small, is also quite irregular in shape. The estimated rest-frame 1500${\rm \ \AA}$ sizes are consistent with the expected sizes of objects at these redshifts as shown in Figure 1 of \citet{ferguson2004} and Figure 7 of \citet{pirzkal2006}. As expected if the size of these objects scale approximately as  ${\rm H^{-1}}$,  the GRAPES LAEs are indeed smaller than what was observed at lower redshifts by \citet{ferguson2004}. On the other hand, we found that the mean ellipticity and concentration of the GRAPES LAEs are both marginally lower than what was measured bt \citet{ferguson2004} and shown in their Figure 3. The mean GRAPES LAE $1500 {\rm \AA}$ ellipticity and concentration are observed to be $0.31 \pm 0.18$ and $2.67 \pm 0.30$, respectively. \\
Morphologically speaking, we observed GRAPES LAEs to be small, concentrated sources with somewhat irregular shapes, reminding us of local dwarf galaxies. The small sizes of the GRAPES LAEs are also similar to the sizes reported for $z\approx 6$ LBGs by \citet{Dow2007}. Several sources (5225, 6139, 9040, 9487) show a resolved clumpy or complex structure ($44\%$ of all sources), while six objects have an effective radius $R_{50} < 1.3 {\rm kpc}$ ($67\%$ of all sources). For purposes of comparison with sources at a lower redshift, \citet{venemans2005} found that, based on 40  $z\approx 3$ sources, $19\%$ of their sample was observed to have clumpy structure, and $25\%$ of their sources had an effective radius $R_{50} < 1.3 {\rm kpc}$. Even considering the relatively low number involved, GRAPES LAEs appear to be systematically smaller, and maybe more irregularly shaped than LAEs at the lower redshift of $z = 3.1$. 

\section{Population Synthesis}\label{popsynth}
The magnitudes shown in Table \ref{phottable} provide us with unprecedented measurements of the rest-frame UV-to-optical light distribution of these faint high-redshift objects. A first glance, Table  \ref{phottable} shows these objects to be brighter in the \mi, \mz, \mJ, and \mH\ bands (corresponding to rest-frame $1200{\rm \ \AA} < \lambda <  2700{\rm \ \AA}$) than in the IRAC bands (optical rest-frame wavelength of $6000{\rm \AA} < \lambda <  7500{\rm \AA}$) and than in the \mb\ and \mv\ bands (rest-frame UV of $700{\rm \AA} < \lambda < 1100{\rm \AA}$). The lack of strong emission in the \mb\ and \mv\ bands is due to HI absorption from the IGM of these objects blue ward of the \lya limit. 

\begin{figure}[h]
\includegraphics[width=7.0in]{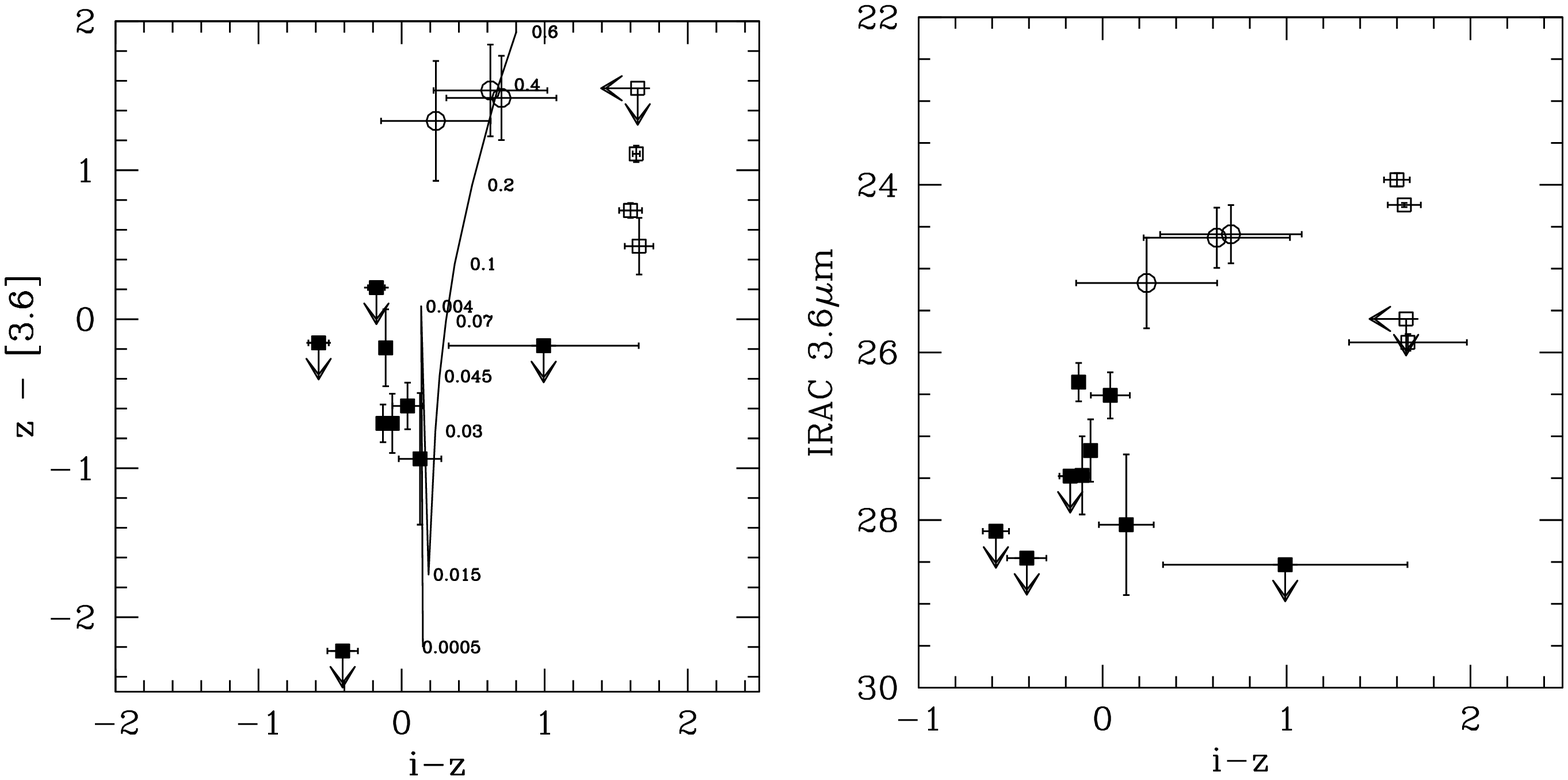}
\caption{\label{colors} Colors of the GRAPES LAEs (filles squares) in z-3.6${\ \micron}.$ vs. i-z space. These objects appear to be extremely blue, similarly to previously identified high-redshift LAEs (circles) from \citet{lai2006}, and significantly bluer than the i-dropout LBGs (open squares) from \citet{eyles2005}. In the left panel,  colors of a z=5 single-burst galaxy are shown (solid line) for comparison with ages (shown in Gyr). This points to the GRAPES LAEs' potentially being very young, low-extinction objects. Similarly, assuming low values of extinction (right), these sources appear to be fainter in the rest-frame optical ($3.6 {\rm {\ \mu}m}$) than either the LBGs or LAEs that were previously observed at these high redshifts.}
\end{figure}

 It is worth emphasizing again that all LAEs were selected spectroscopically and that these sources were not selected using broad band colors. They nevertheless all exhibit very blue intrinsic colors, as shown in Figure \ref{colors}, and appear to be significantly bluer than the LAE samples previously observed by \citet{lai2006} and \citet{eyles2005}. One is immediately drawn to the possibility that these objects might well contain a population of active, young stars, which would of course be consistent with the detection of \lya emission in their GRAPES spectra. There are no known X-ray emitters within 5''  of any of the sources presented here in the Chandra 1Ms observations of the field \citep{giacconi2002}, which excludes the blue nature of these objects from being caused by an active galactic nucleus. confirming earlier conclusions of Malhotra et al. (2003), Wang et al. (2004), Gawiser et al. (2006).
 \\
 A more quantitative determination of the age of these objects, or at least  of the age of the stellar populations of these objects, as well as an estimate of their stellar mass, can be obtained by comparing the spectral energy distribution of these sources with stellar population synthesis models with known star formation history, masses, metallicities, and extinction values.\\
We compared the GRAPES LAEs' SEDs with models from  \citet[][hereafter BC03]{bc03}, using the Paduva evolutionary tracks and using a \citet{salpeter1955} initial mass function (IMF). The choice of a Salpeter IMF allows us to directly compare
GRAPES LAE mass estimates with ones previously published in
the literature.  It also allows us to view our results as upper
limits to the stellar mass of the GRAPES LAEs:  because the
Salpeter IMF continues as an unbroken power law to low stellar
masses, it can overestimate the stellar masses of objects
by a factor of 2 or more if there is a turnover at 
${\rm \sim 1 \ M_\odot}$ \citep[see, e.g.,][]{baldry2003, bell2003, lai2006}. 
The initial goal of this effort, and since many of the magnitudes listed in Table \ref{phottable} are either upper limits or have large associated errors, was to determine whether some reasonable constraints could be set on some of the physical properties of these objects; that is, we might not be in a position to unambiguously determine their exact nature, but we might be able to determine what they are likely {\em not} to be.\\
We used three different set of BC03 models, each with a different SFH: (1) a single instantaneous burst of star formation (SSP), (2) a single but exponentially decaying burst, with an e-folding time of $\tau$ (EXP), and (3) two separate instantaneous star formation bursts separated by an arbitrary amount of time, ${\rm t_{burst}}$, and with each burst contributing an arbitrary fraction of the final stellar mass of the galaxy (2BP). When using either of these three SFHs, the only free parameter that was allowed to vary during the fitting process was the total stellar mass of the model. The redshift, metallicity, age, and extinction were not treated as free parameters but as input assumption of the models.\\
Using each of these three SFH scenarios, we generated an extensive set of simulated spectra. Model spectra were computed at 27 discrete ages, ranging from as low as 5 Myr and up to the age of the Universe at the redshift of interest ($\approx {\rm 1.2 \ Gyr}$; we adopt a $\Lambda$CDM cosmology throughout this paper with $\Omega_M = 0.3$, $\Omega_{\Lambda} = 0.7$, and ${\rm H_0}=70$). The redshifts were fixed to the spectroscopically determined redshifts  listed in Table \ref{objecttable}.
For the SSP and EXP SFH scenarios, we generated model spectra using several  values of the metallicity Z (0.0001, 0.0004, 0.004, 0.008, 0.02 ${\rm [Z_{\sun}]}$, and 0.05) and extinction values \citep[$0 < {\rm A_v} < 3$, using the Calzetti law, ][]{calzetti2000}. The 2BP models assumed a more restricted range of metallicities, using a low metallicity for the first burst (Z=0.0001) and a solar metallicity (Z=0.02) for the second burst, but allowed for the same range of extinction values as the SPP and EXP scenarios. Differential extinction was not included in the 2BP model. We recognize the fact that the two stellar populations could in principle be subject to different amounts of dust extinction. However, GRAPES LAEs are observed to be intrinsically very blue objects and hence to be objects that are dominated by a combination of young stars and low extinction. The data are not sufficient to allow us to study in details whether these two stellar populations do in fact suffer from slightly different amount of dust extinction. IGM absorption \citep{madau1995} was added to all of the model spectra. A total of $\approx 80,000$ separate model spectra were generated at each of the nine redshifts  shown in Table \ref{objecttable}\\

Considering each GRAPES LAE separately, we began the fitting process by minimizing the error-weighted $\chi^2$ per degree of freedom (${\chi}_{\nu}^2$) between observations and a particular model spectrum. During this step, a model spectrum was only allowed to be scaled up or down by an arbitrary amount, thereby  determining the mass of the galaxy assuming that the model spectrum was a good fit to the observations.  We only used bands where an object was detected, and in addition, we excluded  observations in the \mi\ band, which we knew to be contaminated by potentially a large, but uncertain, amount of \lya emission.  We did not attempt to simply correct the broad band \mi\ measurements using the fluxes listed in Table \ref{objecttable}, because of the large uncertainties associated with the aperture correction applied to the measured slitless spectroscopic fluxes. We also excluded the \mb\ and \mv\ bands that we knew to be strongly affected by IGM absorption. \\
This fitting process, while determining a mass estimate for each object, does not however take into account any of the upper limit estimates in cases of a non-detection. We incorporated this additional  information in two distinct ways. First, we computed the Kendall rank correlation value for censored data sets \citep{isobe1986} between individual models (each scaled to best fit bands where the object was detected) and the observations. This method allowed us to quantitatively determine how badly upper limit constraints might be violated by a particular model, but it has the drawback of not being able to take into account the errors in the flux measurements. It has the additional drawback of producing a goodness-of-fit estimator that is strongly quantized because of the limited number of available measurements. While models that strongly violated upper limit constraints could be identified, many of our models led to similar Kendall correlation values, making it difficult to select one model over another. We therefore also relied on another method based on a more generalized version of the error-weighted ${\chi}_{\nu}^2$ described above. This method takes into account upper limit estimates and assigns a ${\chi}_{\nu}^2$ penalty when a model predicts flux values that exceeds established upper limit values (${\chi}_{{\nu}c}^2$). The value of this additive penalty is ${\rm -ln I}$, where  ${\rm I = P(<\sigma| \mu)}$ is the probability that the actual flux is  less than $\sigma$ if the flux is assumed to be normally distributed with mean $\mu$ and width $\sigma$, where $\sigma$ is the flux detection limit of the data and $\mu$ is the flux of the model in that bandpass. This method was used in  \citet{lai2006} where it is described in more details. 

We found a good agreement between the Kendall rank method and the ${\chi}_{{\nu}c}^2$ method, and both methods could indeed be used to reject models that over predicted fluxes in non detected bands. However, and because the  ${\chi}_{{\nu}c}^2$ method allowed us to take flux measurements into account, which was crucial for some  data points (such as the relatively low signal-to-noise ratio ISAAC Ks observations), we eventually relied more on the ${\chi}_{{\nu}c}^2$ method. Models were thus rejected by first computing individual ${\chi}_{{\nu}c}^2$ values and then subsequently determining whether a given ${\chi}_{{\nu}c}^2$ value was an indication that a model was a statistically bad fit to the observations. The exact cut-off value of ${\chi}_{{\nu}c}^2$ used to reject a set of models was determined separately for each object since the number of bands used to compute ${\chi}_{{\nu}c}^2$ differed  from one object to the next.  We defined \Prob\ as the probability of obtaining a value of ${\chi}_{{\nu}c}^2$ as large as the observed ${\chi}_{{{\nu}c}_0}^2$ if the observations really did follow the assumed distribution \citep[see][]{taylor1997}. We then assumed that observations and models differed {\em significantly} when $Prob_d({\chi}_{{\nu}c}^2 > {{\chi}_{{\nu}c_0}^2})$  was less than 5\% ($2 \sigma$ level) and rejected the corresponding model on the basis that it was a poor fit to the observations. Models that could fit the observations could often be identified, using one of the three SFH scenarios that we considered. Some notable exceptions is object 5183, that we found to be somewhat poorly fitted by all of the models that we generated. Cut-off values of \Prob\ were therefore loosened for these two objects, and hence the selected best fits for these objects have higher ${\chi}_{{\nu}c}^2$ values than the rest of the GRAPES LAEs.\\

It is important to stress that we found that properly accounting for the non detection of an object, particularly in the IRAC bands, was crucial when attempting to either include or exclude a particular set of models. The IRAC fluxes, especially the ones that we corrected for the effect of neighbor contamination,  allowed us to determine  an upper  limit to the optical rest frame  flux emitted by the source. We found that such an upper limit imposed a strict upper limit to the number of older stars that could be accounted for in that object. This in turn allowed us to exclude a large subset of our original models. In the following section, we examine the mass estimates for the GRAPES LAEs  and further summarize the range of acceptable input BC03 model parameters.

\section{Mass and Age}\label{ages}
As expected with such a large number of input parameters, we found that a wide range of BC03 models provided an adequate fit to the observations. While a best fitting model could always be identified by simply selecting the model with the lowest value of  ${\chi}_{{\nu}c}^2$ (which we call the best fit model), we also determined the largest mass estimate and the oldest age estimate for each GRAPES LAE by simply examining the subset of BC03 models, and hence the corresponding parameters that went into creating these models, that were not rejected at the $2 \sigma$ level. The model producing the best fit to observations was not always the most massive, or the oldest, model that properly fitted the observations. We defined the maximum-mass models and the maximum-age models as those with the largest mass and age, respectively, that were good fits to the observations (e.g., $2 \ \sigma$ level or higher when possible, or highest possible confidence otherwise).

Table  \ref{bestssp0} summarizes our results using the SSP SFH scenario. This is a  simple model that assumes that all of the stars were created in a single burst event and then simply passively evolved. Examining the best-fit models shown in Table \ref{bestssp0}, we find that all LAEs appear to have very young stellar populations (a few times $10^6 {\rm yr}$) and low stellar masses, with only one object (6139) having a mass estimate that exceeds $10^8 {\rm \ M_{\sun}}$. Extinction values were consistently low, while the metallicities were found to be subsolar in most cases. However, it should be noted that metallicity was not well constrained and that a wide range of metallicities were found to be acceptable. Naturally, a wider range of models were found to properly fit the observations and the masses and ages of the GRAPES LAEs could possibly be somewhat larger than those of the best-fit models, as shown on the last two columns  of Table \ref{bestssp0}. Similarly, a range of age estimates were found to be acceptable. Yet, we found that acceptable ages remained below a few tens of millions years, while masses remained below a few times $\approx 10^8 {\rm \ M_{\sun}}$. Note that none of our BC03 models  successfully fitted object  5183, ot 9040 [\Prob\ $< 5\%$], and that all of our BC03 models could hence formally be rejected at the $2\ \sigma$ level. In these cases, Table \ref{bestssp0} shows the model with the highest possible confidence level.\\

Table \ref{bestexp} is similar to Table \ref{bestssp0} but shows the results obtained using the EXP SFH scenario. In this scenario, star formation occurred in a burst that was not instantaneous but instead exponentially decayed as a function of time, with an e-folding time of $\tau$. Unlike the previous SSP model, not all of the stars have the same age in this scenario, and this scenario allows for an increasingly larger proportion of stars to be older as the galaxy ages. The EXP models would therefore be expected to result in slightly larger age and mass estimates than the SSP scenario. As shown in Table \ref{bestexp}, this is indeed marginally the case. The maximum ages shown in Table \ref{bestexp} are found to be in some cases up to 10 times larger than those produced using the SSP scenario. The maximum masses, in most cases, did not exceed a few times $10^8 {\rm \ M_{\sun}}$ and were only slightly larger than the maximum masses derived using the SSP model.

The mass and ages estimates derived using our third SFH scenario (2BP) are shown in Table \ref{bestssp}. By design, the two-burst scenario was conceived to allow us to fit the SEDs of objects that are intrinsically very blue in their rest-frame UV light yet appear to produce a significant amount of light in the rest-frame optical.  The 2BP scenario allows for the presence of  two distinct stellar populations, one young and blue and another older and therefore relatively redder.  This scenario should therefore provide better fits to observations with both large ACS and IRAC fluxes. It should also be expected to yield higher mass estimates in cases of high upper flux limits in the IRAC bands. This scenario is similar to the \citet{dickinson2003} maximally old population model. As shown in Table \ref{bestssp}, the age estimates derived from the best fit to the data are in some cases significantly larger than those derived using either the SSP or EXP scenario. This is of course achieved by allowing a significant fraction of the stellar population of the galaxy to be in the form of a maximally old ($> 1 {\rm Gyr}$) population of stars. As previously observed, extinction values were low, and we found that lower extinction values were generally preferred using this SFH scenario. But we did find that a range of extinction values and burst mass ratios could fit the observations, leading to larger mass estimates, as long as at a few times $10^6 {\rm \ M_{\sun}}$ of the galaxy mass was in the form of a young stellar population with an age of $\approx (5-35) \times 10^6 {\rm \ yr}$. The younger stellar population accounts for the significant UV flux in these objects, while the much older stars (often up to the age of the universe at the object 's redshift) account for an increase of the total mass of the galaxy by a factor of about 2, when compared with the maximum-mass SSP results. 

\begin{figure}
\includegraphics[width=7.0in]{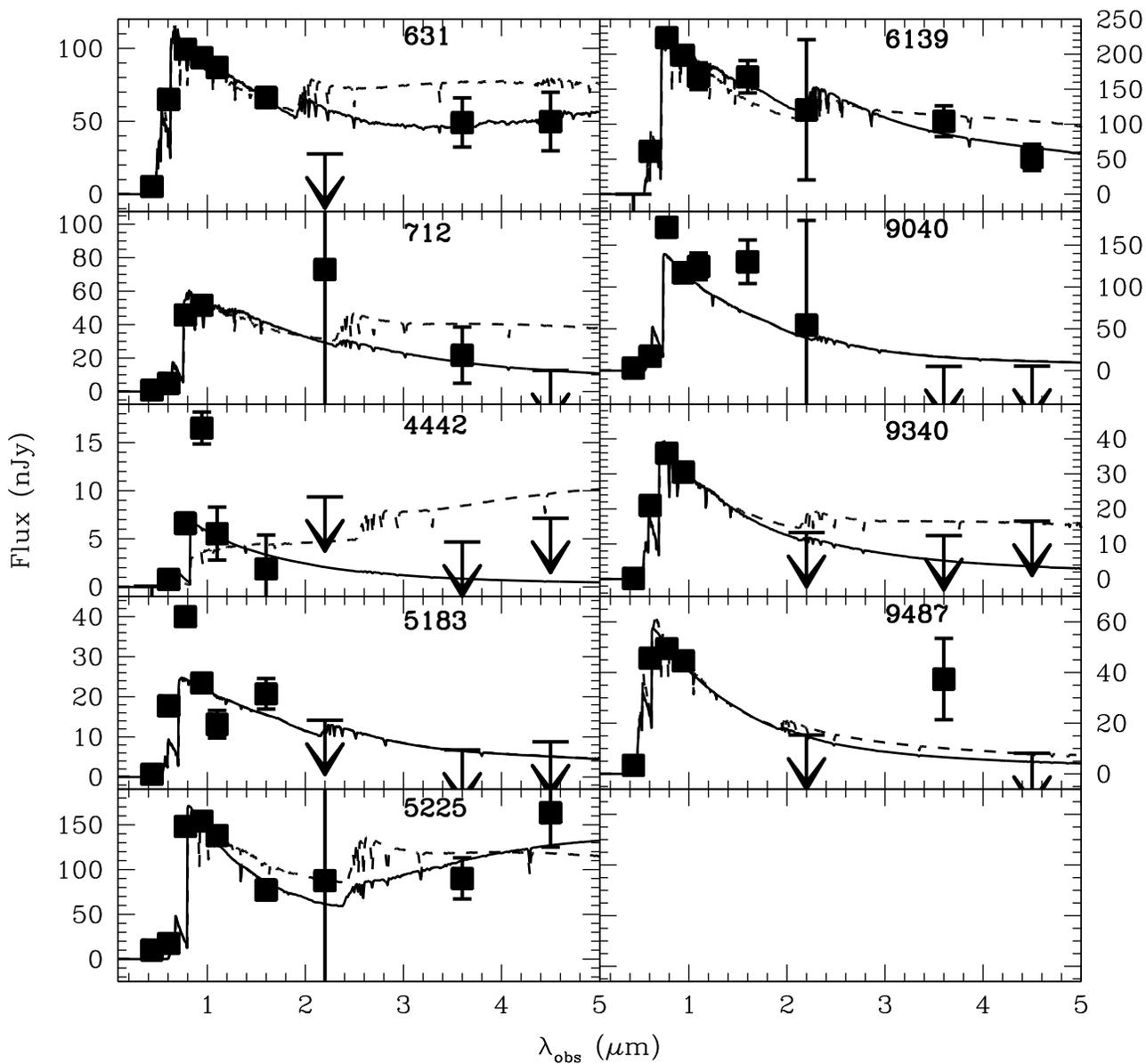}
\caption{\label{plots}Plots of the best double instantaneous burst model (2BP) fits (solid lines) to the observations (squares) and our $1 \ \sigma$ upper limit estimates (arrows). The best, most massive acceptable 2BP models are also shown (dashed lines). The masses, ages, extinctions of each of these models are listed in Table \ref{bestssp}.}
\end{figure}

Figure \ref{plots} shows the best-fit 2BP model for each GRAPES LAE (solid lines), as well as the maximum-mass selected from models that were not rejected at the $2 \ \sigma$ level (dashed line). Note that, as mentioned above, two of the GRAPES LAE sources (5183, and 9040) are not very well fitted by any of our chosen models. It is worth addressing again the possibility that one, or both objects could be a low redshift interloper. However,  and as pointed out in Section \ref{observations}, every single one of these three objects has a very significant color break, consistent with what we would expect from IGM absorption at high redshifts, as well as very blue observed \mz-\mJ\ colors. We performed additional  simulations of the \mb-\mz\ and \mz-\mJ\ colors of these objects that would be expected if they were at much lower (${\rm z \approx 0.25-0.5}$) redshifts, assuming that the detected line was not \lya but [OIII] $\lambda5007$\AA, and that the color break was the result of the  ${\rm 4000\ \AA}$ break. We concluded that the presence of a strong ${\rm 4000\ \AA}$ break, together with a very blue rest-frame near-UV color, cannot be properly accounted for anything but a young stellar population at high redshift, independently of the amount of extinction present, the metallicity of the stellar population, and the age of the stellar population. The latter led us to favor the interpretation that objects 5183 and 9040 are indeed very faint high redshift \lya emitters. We recognize however that future high-resolution spectroscopy will be required in order to unequivocally address this issue.\\

\begin{deluxetable}{lcrcccrr}
\rotate
\tablecaption{Best Single Instantaneous Burst SFH (SSP) Fits to the Observations. \label{bestssp0}}
\tablewidth{0pc}
\tablecomments{Listed are the age, mass, extinction, and metallicity that resulted in the lowest values of  ${\chi}_{{\nu}c}^2$.}
\tablenotetext{a}{Object failed to be properly fitted at the 95\% confidence level}\tabletypesize{\footnotesize} 
\tablehead{
\colhead{UID} & \colhead{Age}  & \colhead{Mass}  & \colhead{Av} & \colhead{Z} & \colhead{\Prob} & \colhead{${\rm Age_{max}}$} & \colhead{${\rm Mass_{max}}$} \\
 \colhead{}  & \colhead{${\rm 10^6 yr}$}   & \colhead{${\rm 10^8 M_{\sun}}$}  & \colhead{} & \colhead{\Zunit}  & \colhead{} & \colhead{${\rm 10^6 yr}$} & \colhead{${\rm 10^8 M_{\sun}}$} 
}
\startdata  
631 & 4.5 & 1.27 & 0.4 & 2.0 & 70.21 & 45.0 & 5.0\\
%  & 9 & 0.312335 & 0.0 & 100.0 & 16.50 & 30.0 & 1.4\\
712 & 3.5 & 0.3 & 0.0 & 100.0 & 93.21 & 45.0 & 13.2\\
%  & 300 & 16.8031 & 0.0 & 20.0 & 92.87 & 800.0 & 73.9\\
4442 & 2.5 & 0.05 & 0.0 & 0.5 & 96.95 & 200.0 & 5.14\\
 % & 1 & 0.0216736 & 0.0 & 20.0 & 3.44 & 1.0 & 0.0217\\
5183\tablenotemark{a} & 2.5 & 0.90 & 0.6 & 2.0 & 3.46 & 2.5 & 0.98\\
 % & 2.5 & 0.166518 & 0.6 & 2.0 & 3.46 & 2.5 & 0.167\\
5225 & 4.5 & 1.34 & 0.0 & 0.5 & 56.30 & 9.0 & 2.89\\
 % & 4.5 & 0.206261 & 0.0 & 0.5 & 56.30 & 9.0 & 0.447\\
6139 & 20. & 10.0  & 0.4 & 0.5 & 78.91 & 30.0 & 29.0\\
 % & 5 & 1.09796 & 0.6 & 0.5 & 86.14 & 150.0 & 22.3\\
9040\tablenotemark{a} & 2.5 & 0.89  & 0.0 & 0.5 & 0.05 & 2.5 & 0.94\\
 % & 5 & 1.01106 & 0.8 & 20.0 & 99.09 & 300.0 & 20.1\\
9340 & 0.5 & 0.16 & 0.0 & 250.0 & 100.00 & 20.0 & 2.57\\
 % & 5 & 0.0384344 & 0.0 & 2.0 & 100.00 & 20.0 & 0.446\\
9487 & 1.0 & 0.40 & 0.05 & 20.0 & 6.46 & 9.0 & 0.82\\
 % & 1 & 0.0792836 & 0.05 & 20.0 & 32.98 & 20.0 & 0.513\\

\enddata
\end{deluxetable}

\begin{deluxetable}{lcrccccrr}
\rotate
\tablecaption{Best Exponentially Decaying Star Burst SFH (EXP) Fits to the Observations \label{bestexp}}
\tablecomments{Listed are the age, mass, extinction, and {\em e}-folding time $\tau$  that resulted in the lowest values of  ${\chi}_{{\nu}c}^2$.}
\tablenotetext{a}{Object failed to be properly fitted at the 95\% confidence level}\tabletypesize{\footnotesize} 

\tablewidth{0pc}
\tabletypesize{\footnotesize} 
\tablehead{
\colhead{UID} & \colhead{Age} & \colhead{Mass}  & \colhead{$\tau$}  & \colhead{Av} & \colhead{Z} & \colhead{\Prob} & \colhead{${\rm Age_{max}}$} & \colhead{${\rm Mass_{max}}$} \\
 \colhead{}  & \colhead{${\rm 10^6 yr}$}   & \colhead{${\rm 10^8 M_{\sun}}$}  & \colhead{Gyr} & \colhead{} & \colhead{\Zunit}& \colhead{}& \colhead{${\rm 10^6 yr}$} & \colhead{${\rm 10^8 M_{\sun}}$} 

}
\startdata  
631 & 8.0 & 0.65 & 0.005 & 0 & 250.0 & 71.06 & 300.0 & 11.\\
 % & 9.0 & 0.124 & 0.0125 & 0 & 250.0 & 13.54 & 150.0 & 1.65\\
712 & 4.5 & 0.30 & 0.005 & 0.0 & 250.0 & 94.09 & 300.0 & 14.8\\
  %& 1200.0 & 44.4 & 0.4 & 0.2 & 100.0 & 97.83 & 1200.0 & 119\\
4442 & 2.5 & 0.07 & 0.001 & 0 & 0.5 & 96.84 & 1200.0 & 6.76\\
  %& 0.5 & 0.0216 & 1 & 0 & 20.0 & 3.44 & 0.5 & 0.0216\\
5183\tablenotemark{a} & 3.0 & 0.81 & 0.005 & 0.6 & 2.0 & 3.38 & 3.0 & 0.81\\
%  & 3.0 & 0.139 & 0.005 & 0.6 & 2.0 & 3.38 & 3.0 & 0.139\\
5225 & 5.0 & 1.41 & 0.001 & 0.05 & 0.5 & 54.41 & 300.0 & 25.3\\
 % & 5.0 & 0.22 & 0.001 & 0.05 & 0.5 & 54.41 & 300.0 & 3.92\\
6139 & 20.  & 18.  & 0.001 & 0.4 & 0.5 & 80.54 & 150.0 & 27.0\\
 % & 800.0 & 12.7 & 1.2 & 0 & 0.5 & 88.68 & 1200.0 & 45.1\\
9040\tablenotemark{a} & 2.5 & 1.2  & 0.001 & 0.0 & 2.0 & 0.03 & 2.5 & 1.24\\
 % & 30.0 & 5.98 & 0.005 & 0.8 & 0.5 & 99.07 & 1200.0 & 34.4\\
9340 & 1.0 & 0.34  & 0.001 & 0.0 & 0.5 & 100.00 & 100.0 & 2.63\\
 % & 20.0 & 0.082 & 1.2 & 0.05 & 0.5 & 100.00 & 100.0 & 0.457\\
9487 & 3.0 & 0.32 & 0.0125 & 0.0 & 20.0 & 6.67 & 15.0 & 0.66\\
 % & 3.0 & 0.0629 & 0.005 & 0 & 20.0 & 33.11 & 45.0 & 0.54\\
 \enddata
\end{deluxetable}

\begin{deluxetable}{lcrccrcrrc}
\rotate
\tablecaption{Best Double Instantaneous Burst SFH (2BP) Fits to  the Observations\label{bestssp}}
\tablecomments{Listed are the age, total mass, and extinction of the object, as well as the age of the second burst and the fraction of the total stellar mass it contains, that resulted in the lowest values of ${\chi}_{{\nu}c}^2$.}\tabletypesize{\footnotesize} 

\tablenotetext{a}{Object failed to be properly fitted at the 95\% confidence level}\tabletypesize{\footnotesize} 
\tablewidth{0pc}
\tabletypesize{\footnotesize} 
\tablehead{
\colhead{UID} & \colhead{Age}  & \colhead{Mass} &  \multicolumn{2}{c}{2nd Burst}    & \colhead{Av} & \colhead{\Prob}   & \colhead{${\rm Age_{max}}$} & \colhead{${\rm Mass_{max}}$} 
 \\
\cline{4-5}
\colhead{} & \colhead{${\rm 10^6 yr}$} & \colhead{${\rm 10^8 M_{\sun}}$} & \colhead{Age (${\rm 10^6 yr}$)} & \colhead{Mass (\%)} & \colhead{} & \colhead{} & \colhead{${\rm 10^6 yr}$} & \colhead{${\rm 10^8 M_{\sun}}$} 
}
\startdata  
631 & 8.5 & 1.56 & 8 & 50  & 0.15 & 73.42 & 1209.0 & 30.9 \\
 % & 10.0 & 0.305 & 9 & 99  & 0 & 16.63 & 1212.5 & 9.06 & 1\\
712 & 16. & 0.68 & 3.0 & 50  & 0.2 & 94.31 & 1010. & 22.7 \\
 % & 1002.5 & 88.6 & 2.5 & 0.5  & 1 & 93.59 & 1200.0 & 129 & 0.325\\
4442 & 2.0 & 0.065 & 1 & 1  & 0 & 96.82 & 870.0 & 11.4  \\
 % & 4.0 & 0.0175 & 0.5 & 1  & 0 & 3.03 & 4.0 & 0.0175 & 1\\
5183\tablenotemark{a} & 17. & 1.09 & 13. & 1  & 0.2 & 3.28 & 17. & 1.08 \\
 % & 16.5 & 0.186 & 12.5 & 1  & 0.2 & 3.28 & 16.5 & 0.186 & 1\\
5225 & 4.5 & 1.28 & 0.5 & 2  & 0.0 & 56.42 & 1000. & 73.8 \\
 % & 4.5 & 0.204 & 0.5 & 2  & 0 & 56.42 & 1001.0 & 11.5 & 2\\
6139 & 20. & 8.70 & 4.5 & 10  & 0.2 & 81.28 & 1006. & 50.38 \\
 % & 1000.5 & 25.5 & 0.5 & 0.875  & 0 & 86.70 & 1045.0 & 59.6 & 0.5\\
9040\tablenotemark{a} & 7.0 & 1.36 & 2.5 & 1  & 0.0 & 0.04 & 7.0 & 1.36 \\
 % & 805.5 & 6.17 & 5.5 & 5  & 0.2 & 98.81 & 1100.0 & 59.9 & 2\\
9340 & 1.5 & 0.18 & 1.0 & 99  & 0.0 & 100.00 & 1200. & 7.47 \\
 % & 14.5 & 0.151 & 9 & 2  & 0.05 & 100.00 & 1203.5 & 1.26 & 3\\
9487 & 1.0 & 0.61 & 0.5 & 8  & 0.15 & 6.30 & 1200. & 1.85 \\
 % & 2.0 & 0.111 & 1 & 2  & 0.2 & 32.93 & 1203.5 & 1.2 & 5\\
\enddata
\end{deluxetable}

\section{Discussion} 

Based on the detection of nine  \lya sources in the HUDF, and using a ${\rm 1/V_{max}}$ methodology, we derive a mean number density of $\approx (1.25 \pm 0.6) \times 10^{-4}  \ \mpcv$. Comparison with previous studies of high redshift \lya sources is difficult,  since each study has slightly different limiting \lya line fluxes and luminosity selection biases. This work in particular does not claim to be a complete sample of LAEs at these redshifts. Still, down to a limiting \lya line luminosity of $2.0 \times 10^{42} {\rm \ ergs\ s^{-1}}$, \citet{taniguchi2005} found that the density of $ z \approx 6.6$ \lya sources is $1.2 \times 10^{-4} \ \mpcv$.  They furthermore estimated, based on observations from \citet{ouchi2003}, that the density of $z=4.9$ \lya sources is  $2 \times 10^{-4} \ \mpcv$, down to the same line luminosity limit. At somewhat lower redshifts, \citet{gawiser2006} recently estimated that the number density (again down the same limiting line luminosity limit ) of $z=3.1$ \lya sources is $(3 \pm 1) \times 10^{-4} \ \mpcv$. If we restrict our sample to the \citet{taniguchi2005} limiting line luminosity, we still detect all 9 sources, and the HUDF LAE density remains $\approx (1.25 \pm 0.6) \times 10^{-4}  \ \mpcv$, which  is consistent with previous works. Similarly, we can also compare  our observations with the results from \citet{rhoads2001} at $z=5.7$. The latter observations were carried out down to the brighter line luminosity limit of $5.3 \times 10^{42} {\rm \ ergs\ s^{-1}}$. In their work, \citet{rhoads2001} inferred a $z=5.7$ \lya density of $\approx 4 \times 10^{-5}  \ \mpcv$. Down to their brighter line luminosity limit, we detect seven LAEs (631, 4442, 5183, 6139, 9040, 9340, and 9487) and derive a \lya density of $(9.68 \pm 4.8) \times 10^{-5}  \ \mpcv$.  Similarly, in this case \citet{taniguchi2005}  computed  a $z \approx 6.6$ \lya object density of $2 \times 10^{-5}  \ \mpcv$, while the $z\approx4.9$ object density from \citet{ouchi2003} becomes $4 \times 10^{-5} \ \mpcv$.  Our observations therefore confirm the relatively low  space  density of \lya objects at these high redshifts, as seen also by \citet{malhotra2004}, confirming an apparent lack of, or at least slow, redshift evolution of the density of LAEs.\\

We found that, no matter which SFR scenario we considered, the GRAPES LAEs were unlikely to be very massive. Both the single instantaneous burst (SSP) and the exponentially decaying burst (EXP) scenarios resulted in similarly low mass estimates, the best EXP models appearing to be those with low values for the $\tau$-parameter. The double-burst scenario (2BP) led to mass estimates that were on average twice that of the other two SFH models. The latter was expected, since the 2BP scenario was conceived to allow fo larger stellar mass estimates by asking the question how large a fraction of the stellar population of a GRAPES LAE could in fact be in the form of an older stellar population, with a lower mass-to-light ratio. 
The different  mass estimates derived using our three distinct SFH models  should be kept in mind when comparing the results presented here with those in the literature. \citet{finkelstein2007} who observed narrow-band-selected \lya sources as part of the Large Area Lyman Alpha (LALA) survey, derived masses for 22 of their sources and estimated them to be in the range $2 \times 10^7$ to $2 \times 10^{9}\ {\rm M_{\sun}}$. They used the best fit of an exponentially decaying burst model (similar to our EXP SFH model). The masses we derived for the GRAPES LAEs using the best-fitting EXP SFH models (Table \ref{bestexp}) are about a factor of 2--3 lower than those found by \citet{finkelstein2007}, with masses varying from $0.1 \times 10^8 $ to $17. \times 10^8 {\rm \ M_{\sun}}$. The largest mass-estimates shown in Table \ref{bestexp} do not exceed $2 \times 10^9 {\rm \ M_{\sun}}$. Using recent observations of  $z\approx 5.7$ \lya sources in the GOODS-N field and a simple SSP model, \citet{lai2006} estimated the masses of their three sources to be in the range $(1.4-5) \times 10^9 {\rm \ M_{\sun}}$, consistent overall  with the results from \citet{finkelstein2007} but, again, significantly larger than most of the best fit Ly-a mass estimates, as well as many of the maximum Ly-a mass estimates derived using the SSP, EXP, or even 2BP SFH model (Tables \ref{bestssp0},\ref{bestexp}, and \ref{bestssp}).

These differences in mass estimate are unlikely to be caused by our respective choices of BC03 templates. Indeed, they are most likely the results of our reach to fainter luminosities (by nearly 2 AB magnitudes) and hence masses. Our sample was not selected based on the presence of a strong broad band continuum emission and hence might constitute a more complete sample (at least in term of masses) of LAEs. As they stand, the GRAPES LAEs  are objects that appear to be  significantly less massive than previously observed LAEs at high redshift, as well as LBGs such as the i-dropout population from  \citet{eyles2005} [with estimated masses of $\approx (1.3-3.8) \times 10^{10} {\rm \ M_{\sun}}$]. This is consistent with these objects being observed to produce most of their light in the rest-frame UV (i.e. a young stellar population with a relatively lower mass-to-light ratio) and to be either faint (e.g., objects 712 and 9487) or even undetected (e.g., objects 4442, 5183, 9040, and 9340) in the rest-frame optical, where one would expect light produced by an older, higher mass-to-light ratio stellar population. \\

We can furthermore examine the star formation rate (SFR) of the HUDF LAEs using either their \lya line luminosities, listed in Table \ref{objecttable}, or their rest-frame UV luminosities \citep{kennicutt1998}. We can additionally use the SFR estimates that the BC03 code produces when generating models. Using the \lya fluxes listed in Table \ref{objecttable}, we derive SFRs that are consistently low for all sources, with an average SFR of $7.8 \pm 3.2 {\rm \ M_{\sun}\ yr^{-1}}$. Similarly, the BC03  SFR estimates imply an equally and systematically low SFR of $3.6 \pm 2.0 {\rm \ M_{\sun}\ yr^{-1}}$. SFR estimates derived using observed broad band rest-frame $2500 {\rm \AA}$ UV are much more uncertain and result in a higher average SFR of  $18.9 \pm 22.9 {\rm \ M_{\sun}\ yr^{-1}}$. However, and independently of which of these SFR estimates we consider, the low mass estimates we derived for the HUDF LAEs imply that they have had ample time to form their stars by the observed redshifts, even with an SFR that is only a few solar masses per year. Strictly speaking, there is no need for the SFR of these objects to have been higher in the past, as was the case for the objects observed by \citet{yan2006}. Higher SFR values would in fact cause these objects to have assembled their stellar masses even more quickly. Based on the observed SFR alone, the HUDF LAEs should have started to form stars no earlier than a few million years before we observed them, unless their SFR was significantly lower in the past, consistent with the fact that we have derived low ages for these objects. 
\\
We are likely to have found very young and very low mass \lya emitters that formed recently as a result of reporting on a whole sample of \lya sources, including objects that were not detected in the rest-frame optical bands. The IRAC observations, when contamination from neighboring objects is properly accounted for,  provide stringent limits to the maximum stellar mass that these objects could have.  It is noteworthy that \citet{gawiser2006} inferred, using a stack of 18 \lya sources at $z \approx 3.1$ \lya, that \lya sources at this redshift should have an average mass that is on the order of $5 \times 10^8 {\rm \ M_{\sun}}$, using models with constant SFH (equivalent to our EXP models with large values of the e-folding time $\tau$). In this paper, we have presented direct observations of such young, low-mass LAEs. 
\\
We conclude that, independently of SFR, metallicities, and extinction, the GRAPES HUDF LAE population is a one of very young objects with low stellar masses of a few million solar masses. There is no easy way to make these objects intrinsically more massive while keeping them intrinsically blue and without violating the IRAC flux estimates and upper limits shown in Table \ref{phottable}. We did derive somewhat larger stellar mass estimates for these objects when we assumed that most of their stars formed at $z >> 10$ in our 2BP models, but the inferred mass of the GRAPES LAEs remained small when compared with previously observed \lya sources.\\

\section{Conclusion}
As part of the GRAPES slitless spectroscopic survey of the HUDF, we have identified a sample of faint LAE galaxies at $4.0 < z < 5.7$. These objects were selected solely based on the detection of \lya emission in their low-resolution ACS slitless spectra. These spectra allowed us to identify LAEs down to extremely faint \lya luminosities ($> 2.3 \times 10^{42} {\rm \ ergs\ s^{-1}}$). The further existence of deep space- and ground-based observations of these objects,  ranging all the way from the rest-frame UV to the rest-frame optical, allowed us to set strong constraints on the masses of these objects. We found that these objects were usually best fitted by models with low extinction and low metallicity. While we attempted to create models using various SFH scenarios, both to compare our results with previous work and to attempt to maximize our mass estimates for these sources, we systematically derived relatively low masses. We conclude that the HUDF LAEs are sources with masses that are significantly lower, by several orders of magnitude, than previously observed LBGs at these redshifts and that they are also significantly less massive than previously observed LAEs at high redshifts. The GRAPES LAEs, with derived star formation rates of $\approx 8 {\rm \ M_\sun \ yr^{-1}}$ and stellar mass estimates that are likely to be no larger than a  few times $10^8 {\rm \ M_{\sun}}$, must have formed quickly, unless their SFR was significantly lower earlier on.  The mean number density of  the GRAPES HUDF LAEs, uncorrected for completeness, is $\approx (1.25 \pm 0.6) \times 10^{-4} \ \mpcv$.  The LAE sources described in this paper are potentially the youngest and least massive galaxies observed to date, and at a time when our universe was just about $\approx 1 {\rm \ Gyr}$ old.

\acknowledgments
This work was supported by grants GO-09793.01-A, GO-10530.11-A and AR-10299.01-A  from the Space Telescope Science Institute, which is operated by AURA under NASA contract NAS5-26555. We would like to thank Haojing Yan for his help identifying potential X-ray counterparts to the HUDF LAEs using the available Chandra data, and Daniel Schaerer, Eric Gawiser, and Barry Rothberg for their helpful comments.

\end{document}